%% file: mkadler.tex
\documentclass{evn2004}
\usepackage{txfonts}
\usepackage{graphicx}
\usepackage{tabularx}
\begin{document}
   \title{Combined VLBI- and X-ray Observations of Active Galactic Nuclei 
          }

   \author{M. Kadler\inst{1}
          \and
	  E. Ros\inst{1}
	  \and
          J. Kerp\inst{2}
	  \and
	  Y.\,Y. Kovalev\inst{3}$^{\rm ,}$\inst{4}
	  \and
	  J.\,A. Zensus\inst{1}
          }

   \institute{Max-Planck-Institut f\"ur Radioastronomie, Auf dem H\"ugel 69, 53121 Bonn, Germany
         \and
              Radioastronomisches Institut der Universit\"at Bonn,
		Auf dem H\"ugel 71, 53121 Bonn, Germany
	 \and
	      National Radio Astronomy Observatory, P.O. Box 2, Green Bank, WV 24944
	 \and
	      Astro Space Center of P.\,N. Lebedev Physical Institute, Profsoyuznaya 84/32, 117997 Moscow, Russia
             }

   \abstract{
Compact radio cores in radio-loud active galactic nuclei (AGNs) are classical targets of 
Very-Long-Baseline interferometric (VLBI) research. Until today little has been known about their 
X-ray properties in comparison to radio-quiet AGNs. Here, we present results from a 
systematic X-ray spectral survey of radio-loud AGNs. This study is based on a statistically 
complete sample of regularly monitored compact, extragalactic radio jets from the 
VLBA 2\,cm Survey and makes use of high-quality X-ray spectroscopic data from the archives of 
the X-ray observatories {\it ASCA}, {\it Beppo}Sax, {\it CHANDRA}, and {\it XMM-Newton}.
Combined VLBI and X-ray observations of AGNs can yield direct 
observational links between the central parsecs of extragalactic jets and dynamical processes 
within the accretion flows around supermassive black holes. It is essential to establish such 
links, which might have the power to attack the unsolved problem: ``What makes an AGN radio-loud?''
We present some early results from our survey project and report briefly on 
detailed studies of 0716+714, NGC\,1052, and 3C\,390.3.
}

   \maketitle
%

\section{Introduction}
The question how powerful relativistic plasma jets are formed in the environment
of supermassive black holes in active galactic nuclei (AGN) is a crucial one. 
Whilst a wealth of information about the jets themselves, in particular about their
variable structure on parsec scales (e.g., Kellermann
et al. \cite{Kel04}), can be obtained from VLBI, 
observational input that may lead to the disclosure of the fundamental difference
between (jet-forming) radio-loud and (jet-suppressing) radio-quiet AGN is rare.
Promising diagnostics 
of jet production are the accretion flows onto the central black
holes. These accretion flows can be probed via X-ray spectroscopy, a technique
which has dramatically benefitted from the advent of the new-generation X-ray
telescopes {\it CHANDRA} and {\it XMM-Newton}. Their broad bandpasses,
combined with the high sensitivity of {\it XMM-Newton} 
and the high angular resolution of {\it CHANDRA}
have greatly improved our ability to identify the 
underlying physical processes of nuclear X-ray emission from AGN. Differences
between the X-ray spectra of radio-loud and radio-quiet AGN may represent different
physical properties of the accretion flows, dilution by an emission component
directly associated with the jet, or even differences in the surrounding medium
of their central engines. 

Compared to radio-quiet AGN, little attention has been 
paid to radio-loud AGN in the X-ray band in the past. Sambruna et al.
(\cite{Sam99}; \cite{Sam02}) have investigated the X-ray spectra of 38 radio-loud AGN
observed by {\it ASCA} and four sources observed by {\it RXTE}. Gambill et al. 
(\cite{Gam03}) analyzed a sample of 17 radio-loud AGN observed with {\it CHANDRA}.
These studies were based on the integrated radio emission measured by single-dish
telescopes or biased towards the presence of prominent extended radio-jet 
structure. To approach the characteristic differences
between both classes of objects, however, it is important to avoid confusion
from considering extended radio emission that arises on kiloparsec scales. In the most
extreme case, emission from extended radio lobes -- millions of light years away from
the central engine -- might still make a galaxy radio-loud when the 
nuclear jet-production has alredy been switched off for ages.

The radio emission of core-dominated radio-loud AGN is heavily
affected by relativistic beaming effects due to large jet velocities  and 
typically small angles to the line of sight. Thus, it is essential to
combine X-ray studies of core-dominated radio-loud AGN with VLBI monitoring 
observations of their parsec-scale radio jets to constrain the jet-system properties.
Ideally, this combination promises to disentangle the
contributions of ``Seyfert-like'' accretion-flow related emission and 
jet-associated (e.g., synchrotron self-Compton) emission in core-dominated 
radio-loud AGN. Ultimately, this approach might allow the fundamental question to
be addressed: ``What makes AGN radio-loud?'' 


\section{An X-ray spectral survey of radio-loud core-dominated AGN}

\input{mkadler_table}

   \begin{figure}[htb]
   \centering
   \includegraphics[width=\columnwidth]{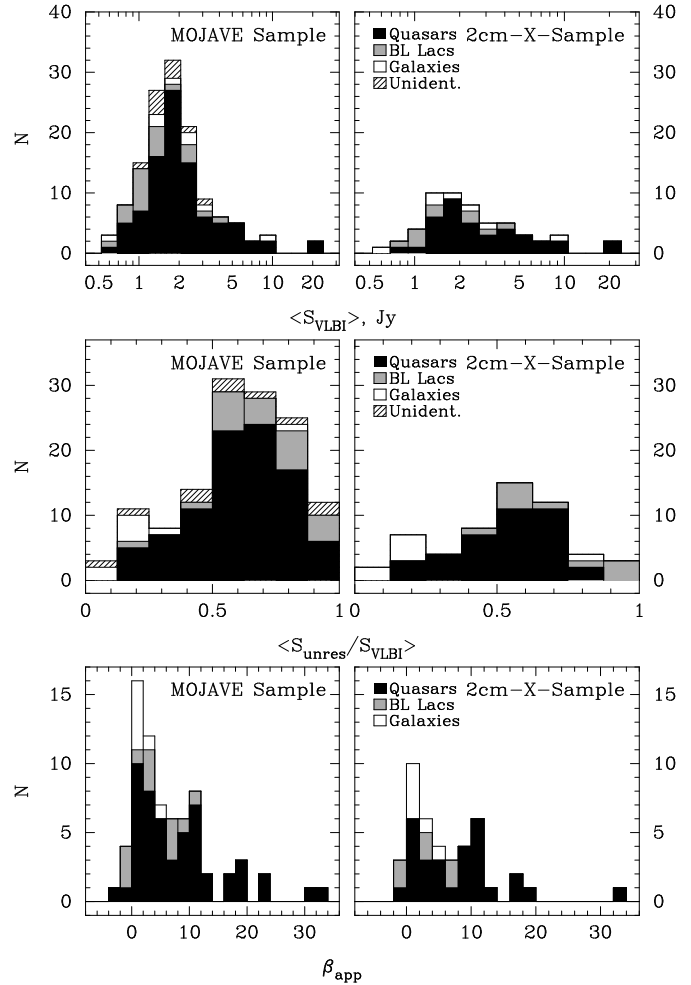}
\vspace{-6pt}
   \caption{\small Distribution of the statistically complete MOJAVE sample and the
2\,cm-X-sample observed at X-rays spectroscopically over the mean total flux density
S$_{\rm VLBI}$ observed by the VLBA (top), the mean compactness index
S$_{\rm unres}$/S$_{\rm VLBI}$ (middle), and the apparent linear velocity 
$\beta = v/c$ of
well determined jet components.
In all histograms the sources are divided by optical
classification into quasars, BL\,Lac objects and galaxies.
\label{fig:histograms}
            }
    \end{figure}

Our sample is based on the statistically complete MOJAVE 
sample 
(e.g.,
Ros \cite{Ros03}; M.\,L.\,Lister et al., in prep.), a Very Long Baseline Array 
(VLBA)
monitored sample of the 133 
brightest core-dominated radio-loud AGN in the northern sky
(see {\tt http://www.physics.purdue.edu/astro/MOJAVE/}).
The kinematics of individual features within
the jet structures of these sources are being monitored since 1994 within the
scope of the VLBA\,2\,cm\,Survey (see e.g., Ros, these proceedings).
We analyze the 
X-ray spectra of radio-loud core-dominated AGN from the MOJAVE sample by making use of
the publicly available archival data
from the telescopes {\it ASCA}, {\it Beppo}Sax,
{\it CHANDRA}, and {\it XMM-Newton}.
Table \ref{tab:sample} gives the 55 resulting sources of our sample                     (the ``2\,cm-X-Sample'' from here on).

Figure~\ref{fig:histograms} (top panels) shows the distribution of mean VLBI radio fluxes
for the 2\,cm-X-Sample in comparison to the distribution for the whole MOJAVE sample. 
In the middle panels of Fig.~\ref{fig:histograms} the comparison of the source
compactness (i.e., unresolved flux on the longest VLBA baselines at 15\,GHz divided
by the total recovered VLBI flux) is shown. These distributions for the whole
MOJAVE sample are analyzed in detail in Y.\,Y.\,Kovalev et al. (in prep.). 
As for the MOJAVE sample,
the peak of the VLBI flux distribution for the 2\,cm-X-Sample reflects the selection
limit while the low flux density wing reflects the variable nature of AGN. The
distribution of the mean source compactness shows, in agreement with expectations
from unification models of radio galaxies, quasars and BL\,Lacs, the 
rising compactness from galaxies to quasars to BL\,Lacs for both samples.
The bottom panels of 
Fig.~\ref{fig:histograms} display the distributions for the 2\,cm-X-Sample
and the whole MOJAVE sample of the apparent linear velocity 
of the
brightest component in each source with well determined kinematics. 
The apparent bimodality of speeds for the full 2\,cm sample and the MOJAVE sample
reported in Kellermann et al. (\cite{Kel04}) is 
present also for the 2\,cm-X-Sample, with a minimum (most pronounced for the quasar
distribution) around $\beta_{\rm app} \sim 8$.

A Kolmogorov-Smirnov test does not reveal any significant differences
between both samples, so that we consider the 2\,cm-X-Sample to be  
representative
of the statistically complete MOJAVE sample. It thus provides a firm basis for 
statistical investigations of core-dominated radio-loud AGN. The statistical analysis
of our sample (e.g., the mean photon index, absorbing column density, luminosity
ratios, and broadband spectral energy distributions) and the comparison to radio-quiet
AGN and the radio-loud AGN samples of Sambruna et al. (\cite{Sam99}; \cite{Sam02})
and Gambill et al. (\cite{Gam03}) will be presented by M.\,Kadler et al. (in prep.).

\begin{figure*}[t!]
\vbox{\includegraphics[clip, width=11cm]{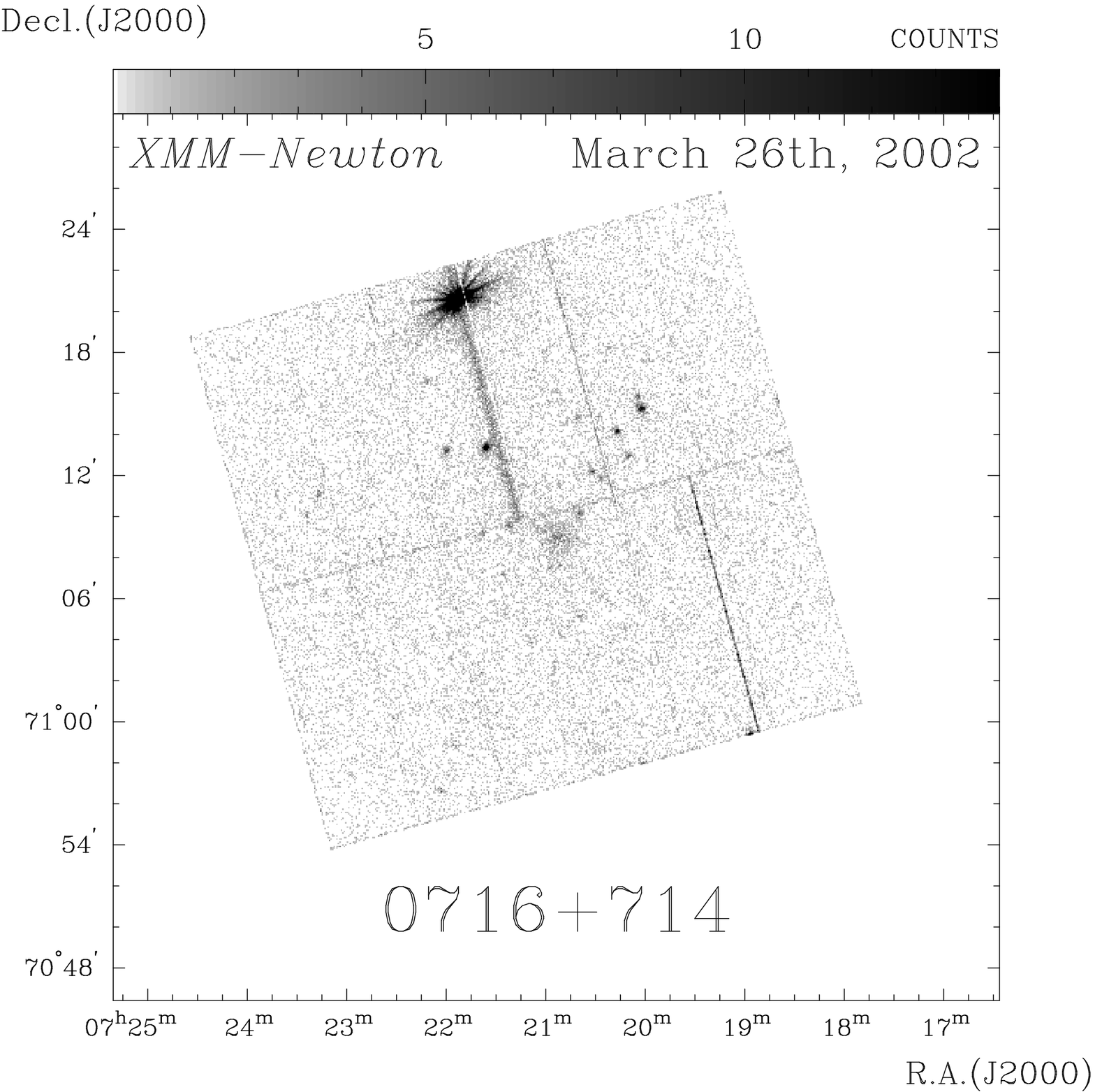}}\vspace{-2.5cm}
\hfill \parbox[b]{6.5cm}{
\caption{\small Raw photon image of the {\it XMM-Newton}
pn-chip during the 6.6\,ksec observation of the galaxy cluster C\,0720.8+7109
in March 2002. 0716+714 is visible as a bright point source close to the top edge of
the image.
\label{fig:0716+714}}
}
\end{figure*}

\section{Radio and X-ray studies of individual sources}
Several individual source studies already have been conducted
connected to this survey project.
We present here three 
cases that are of special interest. Another example is the multi-waveband analysis from
subparsec to megaparsec scales of the 
the superluminal quasar 3C\,454.3, which is presented by 
Pagels et al.\ (these proceedings).
%
\\[6pt]
\noindent
\textbf{\sf 0716+714:}
The BL\,Lac source 
0716+714
was contained inside the field of view of three
{\it XMM-Newton} observations of the galaxy cluster 
C\,0720.8+7109 (see Fig.~\ref{fig:0716+714}).
The X-ray spectral analysis reveals a variable
composition of synchrotron and inverse Compton emission and strong short-term
variability with energy dependent time lags of soft and hard photons (Kadler et al. \cite{Kad04c}). 
So far, 0716+714 has withstood all attempts to determine its redshift directly.
Only indirectly a limit of $z > 0.3$ has been derived
from the non-detection of the optical host galaxy (Wagner et al. \cite{Wag96}).
The {\it XMM-Newton} spectrum shows evidence for line 
emission at $\sim$5.8\,keV in one epoch. If interpreted as iron K$\alpha$ emission
emitted from material at rest in the quasars frame, this implies 
a surprisingly small redshift of 0716+714
of only $z=0.1$. Additional observations with {\it XMM-Newton} and/or {\it Astro-E2}
are necessary to solve this issue. 
\\[4pt]
\noindent
\textbf{\sf NGC\,1052:}
A high angular resolution snapshot {\it CHANDRA} observation of
this low-luminosity AGN revealed the origin of its soft X-ray excess emission.
Soft thermal plasma emission is produced in an extended region around the active
nucleus, most likely originating from jet-driven 
shock heating of the host galaxies interstellar medium (Kadler et al. \cite{Kad04a}).  
The highest signal-to-noise X-ray spectrum of NGC\,1052 comes from a 13\,ksec
{\it XMM-Newton} observation, disclosing the presence of a highly relativistic
broad iron line (Kadler et al. \cite{Kad04b}).
The variable broad iron line and the prominent nuclear radio-jet structure
of NGC\,1052 might provide a direct observational probe of 
jet-disk coupling in an active galaxy.
\\[4pt]
\noindent
\textbf{\sf 3C\,390.3\footnote{\rm 3C\,390.3 is part of the VLBA 2\,cm Survey sample but does not belong to the MOJAVE sample and is not considered for our statistical analysis. We present its X-ray spectrum here because of its relevance for the study of broad iron line emission from radio-loud AGN.}:}
To our knowledge, there is only one source apart from NGC\,1052 for which the profile of a
broad X-ray spectral feature could be significantly better approximated
by a relativistic iron-line model rather than by a narrow or broad Gaussian line 
component: 
the broad-line radio galaxy 3C\,390.3 (Sambruna et al. \cite{Sam99}). 

\begin{figure*}[t!]
\vbox{\includegraphics[clip, width=11cm]{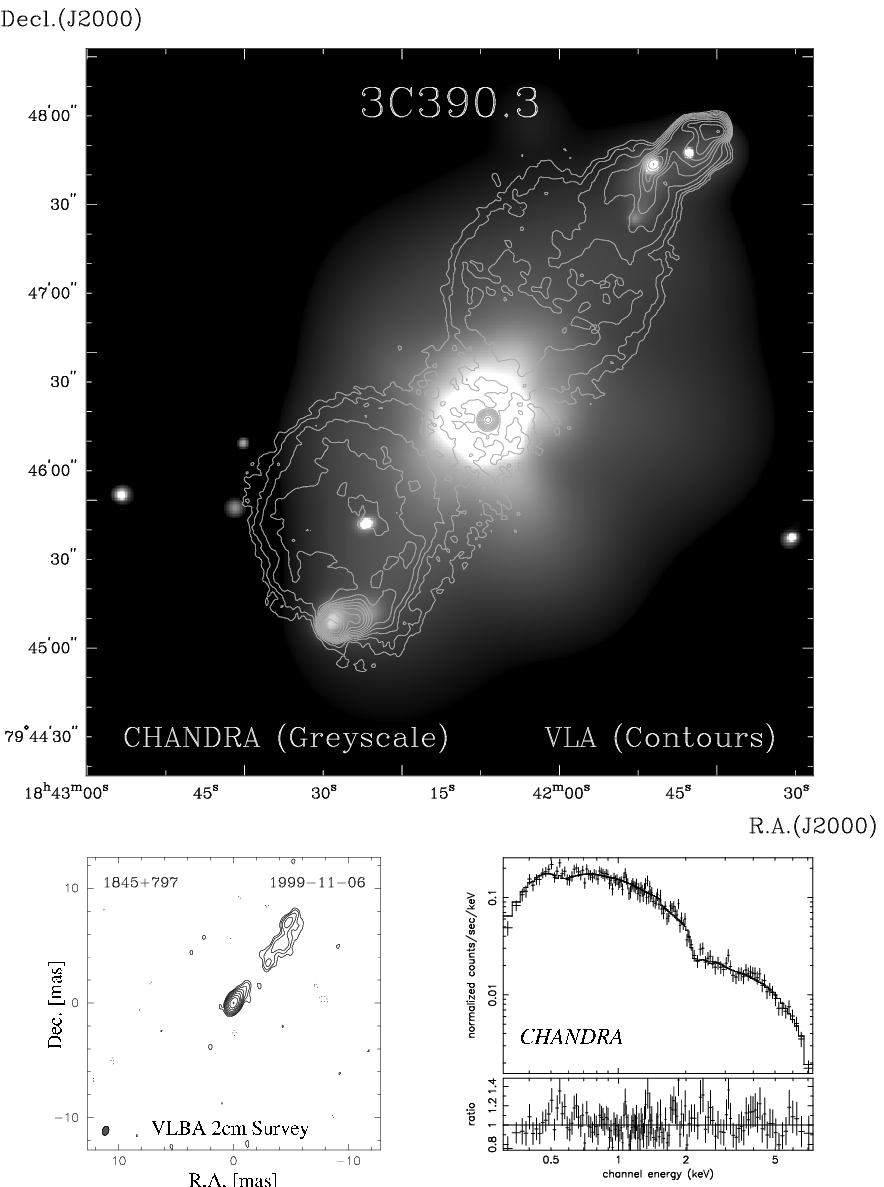}}\vspace{-4.5cm}
\hfill \parbox[b]{6.5cm}{
\caption{\small Top: The kpc-scale radio- and X-ray structure of 3C\,390.3. The 1.6\,GHz
radio jet (taken from the NED archive and previously published by Leahy \& Perley
\cite{Lea91}) is superimposed on the integrated 0.2\,keV to 8\,keV X-ray
brightness distribution from a 35\,ksec {\it CHANDRA} observation (PI: S.\,Wagner).
Bottom left: The milliarcsecond structure of the nuclear radio jet in
November 1999.
Bottom right: X-ray spectrum of 3C\,390.3 and residuals after subtracting a high-energy-reflected power-law
spectrum model.
\label{fig:3c390.3}}
}
\end{figure*}

Here, we report the 
highest signal-to-noise X-ray spectrum of 3C\,390.3 so far obtained, from a 35\,ksec
{\it CHANDRA} observation.
In Fig.~\ref{fig:3c390.3}, we show the X-ray
spectrum of 3C\,390.3 together with its rich large-scale X-ray brightness distribution.
Several knots in the north-west and south-east of the nucleus
can be found, some of them corresponding to knots or hot-spots in the radio structure.
The detailed analysis of these data will be presented elsewhere.
We find a comparably flat
X-ray spectrum ($\Gamma \sim 1.4$ for a simple absorbed-power-law fit), which is best 
approximated by either an absorbed broken power law or a high-energy-reflected power-law
spectrum with an intrinsic photon index of 1.6 to 1.7 in both cases. The residuals
(bottom panel at the top right illustration in Fig.~\ref{fig:3c390.3})
show no obvious evidence for excess iron-line emission. For both models the
fit does not improve significantly when an additional line component is 
added to the model. Thus, the {\it CHANDRA} data do not support the relativistic
iron-line scenario for 3C\,390.3. 



%

   \begin{figure*}[t]
\vbox{\includegraphics[clip,width=12cm]{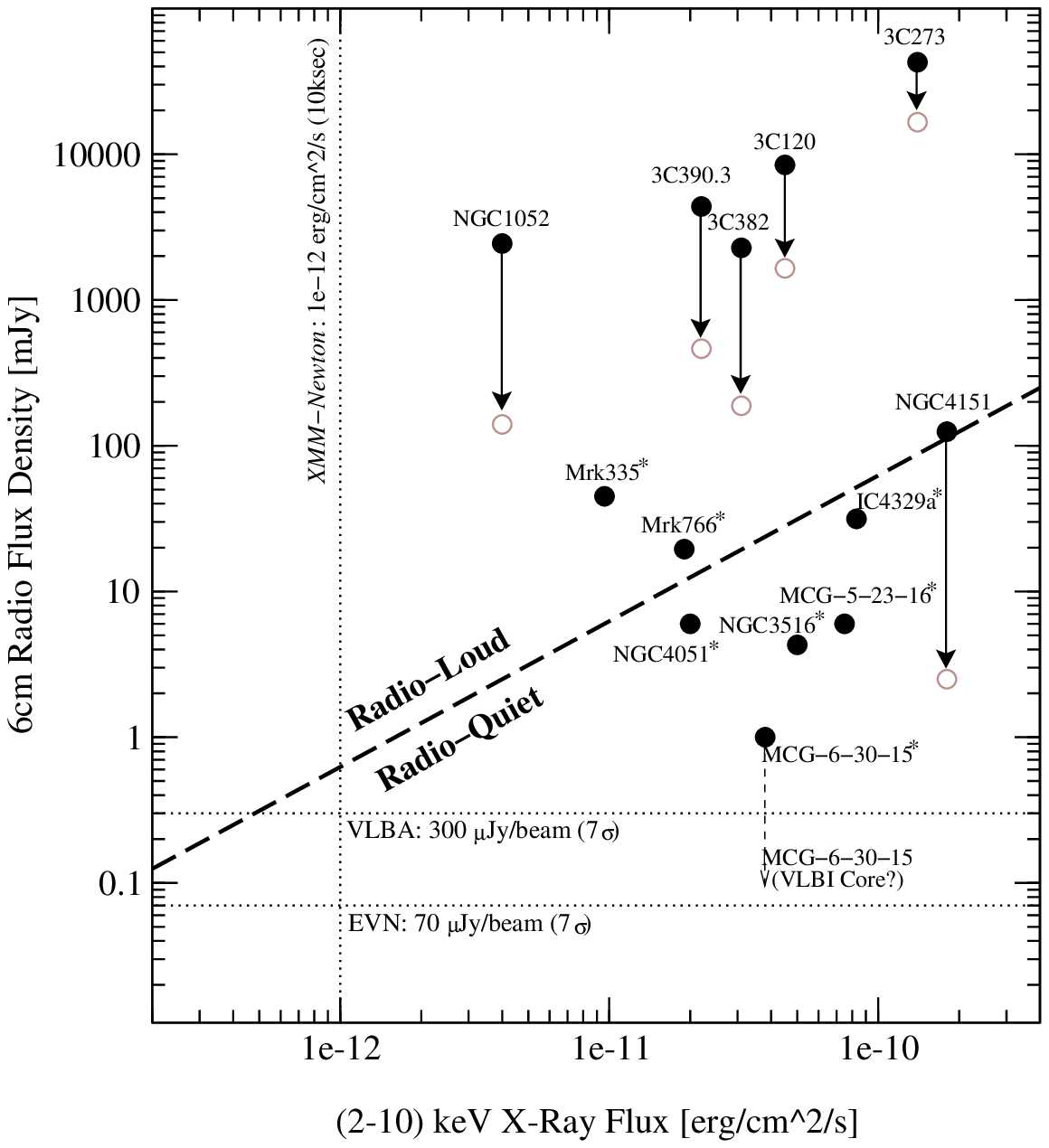}}\vspace{-14.3cm}
\hfill \parbox[b]{4.5cm}{
\vspace{1cm}
   \caption{\small Radio and X-ray fluxes of confirmed and disputed broad-iron-line AGN systems.
The radio-loud/quiet dividing line follows
the definition of Terashima \& Wilson (\cite{Ter03}).
The arrows indicate the range from integrated 
(single-dish: filled circles)
flux density to VLBI-core flux density (open circles).
Only integrated (VLA) flux densities are
given for the radio-quiet objects (marked by an asterisk). 
The arrow for MCG-6-30-15 indicates that the flux density of the
compact core component might be well below the integrated (although compact on VLA scales)
value. Approximate sensitivity
limits for the EVN, the VLBA, and {\it XMM-Newton} are shown. The EVN
and VLBA limits correspond to deep 5 hour integrations with 1024\,Mbps (EVN) and 256\,Mbps
and a 7$\sigma$ detection of an unresolved image feature.
The {\it XMM-Newton} limit at $10^{-12}$\,erg\,cm$^{-2}$\,s$^{-1}$ corresponds to the minimal
source flux for which a 1000-photon spectrum can be obtained in a 10\,ksec pointing.
\label{fig:fluxes}\vspace{1cm}}
            }
    \end{figure*}

\section{Radio-quiet AGN with broad iron lines}
Apart from NGC\,1052, there are a number of confirmed and disputed broad-iron-line detections 
in radio-quiet sources, namely Seyfert\,1 galaxies. Despite their weak radio emission, Seyfert
galaxies often harbor compact radio cores (e.g., Middelberg et al. \cite{Mid04}). 
In some cases, VLBI observations reveal a parsec-scale jet structure in these sources.
Such compact radio cores in radio-quiet 
broad-iron-line systems are potentially attractive targets for high-sensitivity VLBI observations
because of their well studied accretion-disk dynamics.

Figure~\ref{fig:fluxes} displays the positions in the radio-flux-density vs. 
X-ray-flux plane\footnote{Fluxes are taken from 
Fomalont et al. (\cite{Fom00}),
Giovannini et al. (\cite{Gio01}),
Gregory \& Condon (\cite{Gre91}),
Morganti et al. (\cite{Mor99}),
Rush et al. (\cite{Rus96}),
Sambruna et al. (\cite{Sam99}),
Stickel et al. (\cite{Sti94}),
Ulvestad \& Wilson (\cite{Ulv84}), 
and the HEASARC website.}
of the undisputed broad-iron-line 
Seyfert galaxies MCG-6-30-15 (Fabian et al. \cite{Fab02}; Wilms et al. \cite{Wil01}), 
MCG-5-23-16 (Dewangan et al. \cite{Dew03}), NGC\,3516 (Turner et al. \cite{Tur02}), 
Mrk\,335 (Gondoin et al. \cite{Gon02}), and
Mrk\,766 (Pounds et al. \cite{Pou03}). In addition, 
the sources IC\,4329a (Gondoin et al. \cite{Gon01}; Mc\,Kernan \& Yaqoob \cite{McK04}), 
and NGC\,4151 (Schurch et al. \cite{Sch03}) are shown, in which 
{\it ASCA} detected broad iron lines which could not be confirmed by {\it XMM-Newton}
or are disputed.
Finally, the radio-loud sources 3C\,120 (Ogle et al. \cite{Ogl04}), 
3C\,273 (Page et al. \cite{Pag04}; Yaqoob \& Serlemitsos \cite{Yaq00}), 
3C\,382 (Sambruna et al. \cite{Sam99}), 
3C\,390.3 (Sambruna et al. \cite{Sam99} and this work), and NGC\,1052 (Kadler et al. \cite{Kad04b}) are also plotted.
Figure~\ref{fig:fluxes} shows that not only the radio-loud AGN are accessible by 
VLBI, but also the so-called radio-quiet Seyfert galaxies. In particular, the EVN
with its large antennas provides enough sensitivity to image, e.g., the parsec-scale
structure of the possibly most interesting broad-iron-line Seyfert galaxy
MCG-6-30-15, given that it harbors a compact nucleus on milliarcsecond scales. 
 

\begin{acknowledgements}
M.\,K. was supported for this research through a stipend from the International
Max Planck Research School (IMPRS) for Radio and Infrared Astronomy at the University of
Bonn. 
Y.\,Y.\,K. is a Jansky Fellow.
Part of this work was done within the framework of the VLBA\,2cm\,Survey collaboration (see {\tt http://www.cv.nrao.edu/2cmSurvey}). This research was supported by the European Commission's I3 Programme
``RADIONET", under contract No.\ 505818.
{The VLBA
is an instrument of the U.S.A. National Radio Astronomy Observatory,
which is a facility of the National Science Foundation, operated under cooperative
agreement by Associated Universities, Inc.} 
{X-ray archival data are made available from the High Energy Astrophysics Science Archive                                                    Research Center (HEASARC), provided by NASA's Goddard Space Flight Center.} 
\end{acknowledgements}

\end{document}

%% file: mkadler_table.tex
\begin{table}[htb]
\caption{\small The 2cm-X-Sample}
\label{tab:sample}
\[
\resizebox{\columnwidth}{!}{%
\centering
\begin{tabular}{@{}c@{~\,}c@{~\,}c@{~\,}c@{~\,}c@{~\,}c@{~\,}c@{~\quad}c@{~\,}c@{~\,}c@{~\,}c@{~\,}c@{~\,}c@{~\,}@{}}
\cline{1-6}\cline{8-13}
\noalign{\smallskip}
\cline{1-6}\cline{8-13} 
\noalign{\smallskip}
Source$^{\rm a}$   & Alt. Name &{\it C}$^{\rm 1}$&{\it X}$^{\rm 2}$&{\it A}$^{\rm 3}$&{\it B}$^{\rm 4}$&&Source$^{\rm a}$   & Alt. Name &{\it C}$^{\rm 1}$&{\it X}$^{\rm 2}$&{\it A}$^{\rm 3}$&{\it B}$^{\rm 4}$\\
\noalign{\smallskip}
\cline{1-6}\cline{8-13}
\noalign{\smallskip}  
0007+106 & III\,Zw\,2&  & $\surd$ & $\surd$ & $\surd$ && 1228+126 & M\,87     &$\surd$ & $\surd$ & $\surd$ & $\surd$   \\
0048$-$097 & OB\,$-$080&   &   &   & $\surd$ &&1253$-$055 & 3C\,279   &$\surd$ &   & $\surd$ & $\surd$   \\
0202+149 & 4C\,15.05 &   &   & $\surd$ &   &&1308+326 & OP\,+313  &  & $\surd$ & $\surd$ &     \\
0234+285 & CTD\,20   &   &   &   & $\surd$ &&1334$-$127 & OP\,$-$158.3&  & $\surd$ &   &     \\
0235+164 & OD\,+160  & $\surd$ & $\surd$ & $\surd$ & $\surd$ &&1413+135 & OQ\,+122  &  &   & $\surd$ &     \\
0238$-$084 & NGC\,1052 & $\surd$ & $\surd$ & $\surd$ & $\surd$ &&1458+718 & 3C\,309.1 &$\surd$ &   &   &     \\
0316+413 & 3C\,84    & $\surd$ & $\surd$ & $\surd$ & $\surd$ &&1502+106 & 4C\,10.39 &  & $\surd$ &   &     \\
0333+321 & NRAO\,140 &   &   & $\surd$ & $\surd$ &&1510$-$089 & OR\,$-$017  &$\surd$ &   & $\surd$ & $\surd$   \\
0415+379 & 3C\,111   &   & $\surd$ & $\surd$ & $\surd$ &&1611+343 & DA\,406   &  &   &   & $\surd$   \\
0420$-$014 & OA\,+129  &   &   & $\surd$ & $\surd$ &&1633+382 & 4C\,38.41 &  &   & $\surd$ &     \\
0430+052 & 3C\,120   & $\surd$ & $\surd$ & $\surd$ & $\surd$ &&1641+399 & 3C\,345   &$\surd$ &   &   & $\surd$   \\
0458$-$020 & DA\,157   & $\surd$ &   &   &   &&1655+077 & OS\,+092  &$\surd$ &   &   &     \\
0528+134 & OG\,+147  &   &   & $\surd$ & $\surd$ &&1741$-$038 & OT\,$-$068  &  &   &   & $\surd$   \\
0605$-$085 & OH\,$-$010  & $\surd$ &   &   &   &&1749+096 & 4C\,09.57 &  &   & $\surd$ &     \\
0716+714 &           &   & $\surd$ & $\surd$ & $\surd$ &&1803+784 &           &  &   &   & $\surd$   \\
0735+178 & DA\,237   &   &   & $\surd$ &   &&1823+568 & 4C\,56.27 &  &   &   & $\surd$   \\
0736+017 &           &   &   & $\surd$ & $\surd$ &&1828+487 & 3C\,380   &$\surd$ &   &   &     \\
0738+313 & OI\,+363  & $\surd$ & $\surd$ &   &   &&1928+738 & 4C\,73.18 &$\surd$ &   & $\surd$ & $\surd$   \\
0827+243 &           & $\surd$ &   &   &   &&1936$-$155 &           &  &   &   & $\surd$   \\
0836+710 & 4C\,71.07 & $\surd$ & $\surd$ & $\surd$ & $\surd$ &&1957+405 & Cyg\,A    &$\surd$ &   & $\surd$ & $\surd$   \\
0851+202 & OJ\,+287  &   &   & $\surd$ & $\surd$ &&2134+004 & DA\,553   &  &   &   & $\surd$   \\
0923+392 & 4C\,39.25 & $\surd$ &   & $\surd$ &   &&2145+067 &           &  &   & $\surd$ &     \\
1038+064 & 4C\,06.41 &   & $\surd$ & $\surd$ &   &&2200+420 & BL\,Lac   &$\surd$ &   & $\surd$ & $\surd$   \\
1055+018 & 4C\,01.28 & $\surd$ &   &   &   &&2223$-$052 & 3C\,446   &  &   & $\surd$ & $\surd$   \\
1127$-$145 & OM\,$-$146  & $\surd$ & $\surd$ &   & $\surd$ &&2230+114 & CTA\,102  &  &   & $\surd$ & $\surd$   \\
1156+295 & 4C\,29.45 & $\surd$ &   &   &   &&2243$-$123 & OY\,$-$176  &  &   &   & $\surd$   \\
1222+216 &           & $\surd$ &   & $\surd$ &   &&2251+158 & 3C\,454.3 &$\surd$ &   &   & $\surd$   \\
1226+023 & 3C\,273   & $\surd$ & $\surd$ & $\surd$ & $\surd$ &         &           &                \\
\noalign{\smallskip}
\cline{1-6}\cline{8-13}
\end{tabular}
}
\]
{\scriptsize $^{\rm a}$\,B\,1950.0 coordinates; $^{\rm 1}$\,Archival {\it CHANDRA} data; $^{\rm 2}$\,Archival {\it XMM-Newton} data; $^{\rm 3}$\,Archival {\it ASCA} data; $^{\rm 4}$\,Archival {\it Beppo}Sax data 
}
\end{table}

%% file: mkadler.bbl
\begin{thebibliography}{}
\small

\bibitem[2003]{Dew03} Dewangan, G.\,C., Griffiths, R.\,E., \& Schurch, N.\,J.\ 2003, ApJ, 592, 52
\bibitem[2002]{Fab02} Fabian, A.\,C., Ballantyne, D.\,R., Merloni, A., et al.\ 2002, MNRAS, 331, L35
\bibitem[2000]{Fom00} Fomalont, E.~B., Frey, S., Paragi, Z., et al. \ 2000, ApJS, 131, 95
\bibitem[2003]{Gam03} Gambill, J.~K., Sambruna, R.\,M.\, Chartas, G., et al. 2003, A\&A, 401, 505
\bibitem[2001]{Gio01} Giovannini, G., Cotton, W.~D., Feretti, L., Lara, L., \& Venturi, T.\ 2001, ApJ, 552, 508
\bibitem[2001]{Gon01} Gondoin, P.,  Barr, P., Lumb, D., et al. 2001, A\&A, 378, 806
\bibitem[2002]{Gon02} Gondoin, P., Orr, A., Lumb, D., \& Santos-Lle\'o, M.\ 2002, A\&A, 388, 74
\bibitem[1991]{Gre91} Gregory, P.\,C., \& Condon, J.\,J. 1991, ApJS, 75, 1011  
\bibitem[2004a]{Kad04a} Kadler, M., Kerp, J., Ros, E., et al. 2004a, A\&A, 420, 467 
\bibitem[2004b]{Kad04b} Kadler, M., Ros, E., Weaver, K.\,A., Kerp, J., Zensus, J.\,A. 2004b, BAAS 36, No.\,2, 823  
\bibitem[2004c]{Kad04c} Kadler, M., Kerp, J., \& Krichbaum, T.\,P. 2004c, A\&A, submitted 
\bibitem[1989]{Kel89} Kellermann, K.\,I., Sramek, R., Schmidt, M., Shaffer, D.\,B., \& Green, R.\ 1989, ApJ, 98, 1195
\bibitem[2004]{Kel04} Kellermann, K. I., Lister, M.\,L., Homan, D.\,C., et al. 2004, ApJ, 609, 539
\bibitem[1991]{Lea91} Leahy, J.~P.~\& Perley, R.~A.\ 1991, ApJ, 102, 537
\bibitem[2004]{McK04} McKernan, B.~\& Yaqoob, T.\ 2004, ApJ, 608, 157
\bibitem[2004]{Mid04} Middelberg, E., Roy, A.\,L., Nagar, N.\,M., et al. 2004, A\&A , 417, 925
\bibitem[1999]{Mor99} Morganti, R., Tsvetanov, Z.\,I., Gallimore, J., \& Allen, M.\,G. 1999, A\&AS, 137, 457
\bibitem[2004]{Ogl04} Ogle, P.\,M., Davis, S.\,W., Antonucci, R.\,R.\,J., et al. 2004, BAAS 36, No.\,2, 766
\bibitem[2004]{Pag04} Page, K.\,L., Turner, M.\,J.\,L., Done, C., et al. 2004, MNRAS, 349, 57 
\bibitem[2003]{Pou03} Pounds, K.\,A., Reeves, J.\,N., Page, K.\,L., Wynn, G.\,A., \& O'Brien, P.\,T.\ 2003, MNRAS, 342, 1147
\bibitem[2003]{Ros03} Ros, E. 2003, in Highlights in Spanish Astrophysics III, J. Gallego, J. Zamorano, N. Cardiel (eds.), Kluwer Academic Publishers, p.235
\bibitem[1996]{Rus96} Rush, B., Malkan, M.\,A., \& Edelson, R.\,A. 1996, ApJ, 473, 130
\bibitem[1999]{Sam99} Sambruna, R.\,M., Eracleous, M., \& Mushotzky, R.\,F. 1999, ApJ, 526, 60 
\bibitem[2002]{Sam02} Sambruna, R.\,M., Eracleous, M., \& Mushotzky, R.\,F. 2002, New Astron. Rev., 46, 215 
\bibitem[2003]{Sch03} Schurch, N.\,J., Warwick, R.\,S., Griffiths, R.\,E., \& Sembay, S.\ 2003, MNRAS, 345, 423
\bibitem[1994]{Sti94} Stickel, M., Meisenheimer, K., \& K\"uhr, H. 1994, A\&AS, 105, 211
\bibitem[2003]{Ter03} Terashima, Y., \& Wilson, A.\,S. 2003, ApJ, 583, 145
\bibitem[2002]{Tur02} Turner, T.\,J., Mushotzky, R.\,F., Yaqoob, T. et al.\ 2002, ApJ, 574, L123
\bibitem[1984]{Ulv84} Ulvestad, J.\,S., \& Wilson, A.\,S. 1984, ApJ, 285, 439
\bibitem[1996]{Wag96} Wagner, S.\,J., Witzel, A., Heidt, J., et al. 1996, AJ, 111, 2187
\bibitem[2001]{Wil01} Wilms, J., Reynolds, C.\,S., Begelman, M.\,C., et al. 2001, MNRAS, 328, L27
\bibitem[2000]{Yaq00} Yaqoob, T.,~\& Serlemitsos, P.\ 2000, ApJ, 544, L95
\end{thebibliography}
